\documentclass[reprint,amsmath,amssymb, aps,prb,superscriptaddress]{revtex4-2}

\usepackage[english]{babel}
\usepackage{graphicx}
\usepackage{dcolumn}
\usepackage{bm}
\usepackage{physics}
\usepackage{placeins}
\usepackage{soul}
\usepackage{hyperref}
\usepackage[dvipsnames]{xcolor}

\hypersetup{
    colorlinks,
    linkcolor={red!40!black!100!},
    citecolor={blue!30!black!100},
    urlcolor={blue!30!black!100}
}

\begin{document}
\selectlanguage{english}

\preprint{ARK,SR,UniLu}

\title{
Flipping of electronic spins in BiFeO$_3$ via chiral $d-d$ excitations}

\author{Aseem Rajan Kshirsagar}
\affiliation{%
Department of Physics and Materials Science, University of Luxembourg, Luxembourg}%

\affiliation{%
Univ. Rennes, ENSCR, INSA, CNRS, Institut des Sciences Chimique de Rennes (UMR6226), France}%

\email{aseem@posteo.net, aseem.kshirsagar@uni.lu}

\author{Sven Reichardt}%
 
\affiliation{%
Department of Physics and Materials Science, University of Luxembourg, Luxembourg}%

\date{\today}

\begin{abstract}
BiFeO$_3$ is a multiferroic material featuring ferroelectricity and noncollinear antiferromagnetism.
Definitive and efficient control of the characteristic spin texture of BiFeO$_3$ is attractive for emerging quantum devices.
In this regard, crystal-field $d\rightarrow d$ excitations localized on Fe atomic sites in BiFeO$_3$ provide an avenue for manipulation of the spin texture as they induce a complex interplay among the spin, charge, and lattice degrees of freedom.
In this work, the \textit{ab initio} \textit{GW}-BSE method is used to characterize these excitations within an excitonic picture.
We find that the $d-d$ transitions appear as strongly bound, chiral, spin-flip excitons deep within the electronic band gap as a result of the intricate competition between the lattice potential, the antiferromagnetic ordering, the spin-orbit coupling, and the electron-hole interaction.
Most crucially, these excitons are composed of electron-hole pairs with opposite spins that constitute almost all of their $\pm \hbar$ total angular momentum.
These excitons of specific angular momentum can be selectively excited using circularly polarized light, consequently modulating the local magnetic moment. 
\end{abstract}

\maketitle

Multiferroics are materials which exhibit multiple ferroic degrees of freedom~\cite{Fiebig2016-sx}.
In these materials, the coupling between different ferroic properties, \textit{e.g.}, between ferromagnetism and ferroelectricity, can be exploited to effectively manipulate one ferroic property by tuning another.
Among this class of materials, bismuth ferrite (BiFeO$_3$) stands out as an extensively studied, single-phase, room-temperature multiferroic~\cite{Kuo2016}.
It features a magnetic order in the form of a spin cycloid of period $\sim 62$~nm ~\cite{Sosnowska1982,Ratcliff2016-nf,Xu2021-wu}, caused by the canting of the antiferromagnetically aligned Fe spins due to an antisymmetric exchange (Dzyaloshinskii–Moriya) interaction~\cite{AFM_neel_1,AFM_neel_2,Ederer2005-nf,Meyer2022-ab}.
At the same time, BiFeO$_3$ possesses a large intrinsic electric polarization of $\sim 90\mu$C/cm$^2$, due to a structural distortion along the [111] crystal direction attributed to a repulsion between the Fe and Bi valence electrons~\cite{Moreau1971,Catalan2009,wang_epitaxial_2003,Spaldin2017,Seshadri2001,Neaton2005}.
The coupling between these ferroelectric and magnetic degrees of freedoms has been exploited to, \textit{e.g.}, achieve electrostatic control over magnetic domains~\cite{Heron2014-qa,Haykal2020,Zhao2006} and over spin waves or magnons ~\cite{deSousa2008,Rovillain2010-kb,Sando2013,Parsonnet2022}.
This control makes BiFeO$_3$ an attractive physical platform to design magnetoelectric and spintronic devices on~\cite{Manipatruni2019,deSousa2022}.

Beyond the mentioned electrostatic control, several other studies have also explored the coupling of spins in BiFeO$_3$ with optical excitations within the electronic band gap~\cite{xu_optical_2009,ramirez_magnon_2009,Ramirez2009-rx,Galuza1998,Wei2017-bd}.
These excitations have been identified as localized transitions within the Fe $3d$-shell ($d$-$d$), which involve the flipping of the electron spin.
These excitations have been linked to photo-induced in-gap features observed in recent optical transmission experiments ~\cite{meggle_2019,meggle_optical_2019,Burkert2016}.
They also couple strongly to the nuclear degrees of freedom, as evidenced by their prominent appearance in resonant Raman scattering intensities~\cite{Cazayous2009-go,weber_temperature_2016,Ramirez2009-rx,Singh2008-va}.
These excitations thus provide a very promising means to optically control the charge, spin, and vibrational degrees of freedom in BiFeO$_3$, which can be attractive for a variety of applications, ranging from sensing to quantum information devices.
However, in order to effectively exploit these electronic excitations and leverage their potential, their thorough characterization is necessary. \textcolor{black}{Ab initio many-body methods have been recently employed to study analogous $d-d$ transitions in several crystalline transition metal compounds~\cite{Acharya2022,Acharya2023,wu_physical_2019,Wu2022-kg}.  }  \par 

In this Letter, as well as in our companion paper~\cite{aseem_sven_prb}, we present a detailed theoretical characterization of the in-gap, spin-flip excitations of BiFeO$_3$ using state-of-the-art density functional and many-body perturbation theory methods.
We find that the half-filled spin configuration of the Fe $3d$-shell imposes constraints on the electronic excitation spectrum, which lead to a localization of the  spin-flip excitation on a single Fe site, while spin-conserving excitons are delocalized across different Fe centres.
The delicate interplay of the symmetry-based crystal field spitting of the Fe-$d$ shell, the electron-hole attraction, and the spin-orbit coupling (SOC) yields a unique fine structure of the exciton spectrum of BiFeO$_3$, which has not been well understood so far.
The crystal symmetry, lacking an inversion center, 
in combination with the SOC, leads to spin-polarized exciton states which exclusively couple to left or right circularly-polarized light (CPL), opening up possibilities for tuning chiral magnetic excitations of BiFeO$_3$ with light\textcolor{black}{, as demonstrated recently for Fe$_2$Mo$_3$O$_8$~\cite{Sheu2019-kk,Zhuang2023-hm}.}

We use density functional theory (DFT) in the formulation of Kohn and Sham~\cite{ks-dft} to obtain the ground-state electronic properties of BiFeO$_3$.
The exchange and correlation potential is approximated with the semi-local PBE functional~\cite{Perdew1996}.
The systematically underestimated PBE electronic transition energies are corrected using the eigenvalue self-consistent \textit{GW} ($evGW$) method~\cite{Hedin1965-sv,Onida2002-qg,Hybertsen1985-ne,Shishkin2007-hj}.
For the description of the bound electron-hole pairs, we construct and solve the Bethe-Salpeter equation (BSE) in the basis of electronic transitions from occupied to unoccupied DFT orbitals~\cite{Salpeter1951-va,Rohlfing2000-sm,Strinati1982-zz,Palummo2004-xk}.
The interaction between the hole and electron is approximated as the sum of an attractive, statically screened Coulomb interaction and the repulsive, unscreened exchange interaction.
Solving of the BSE yields the exciton energies $E_S$ and exciton envelope wave functions $A^S_{\mathbf{k},c,v} = \langle \mathbf{k},v \to \mathbf{k},c | S \rangle$, which represent the overlap of the exciton state $|S\rangle$ with the transition from an occupied, valence state $|\mathbf{k},v\rangle$ to an empty, conduction band state $|\mathbf{k},c\rangle$, where $\mathbf{k}$ denotes a wave vector in the first Brillouin zone.
A complete and thorough description of the computational methods can be found in our companion work~\cite{aseem_sven_prb} and references therein~\cite{Giannozzi2009,Andreussi2017-tq,Marini2009,Sangalli2019,Hamann2013,pseudodojo_2018}.

\begin{figure*}[]
    \centering
    \includegraphics[width=0.85\textwidth]{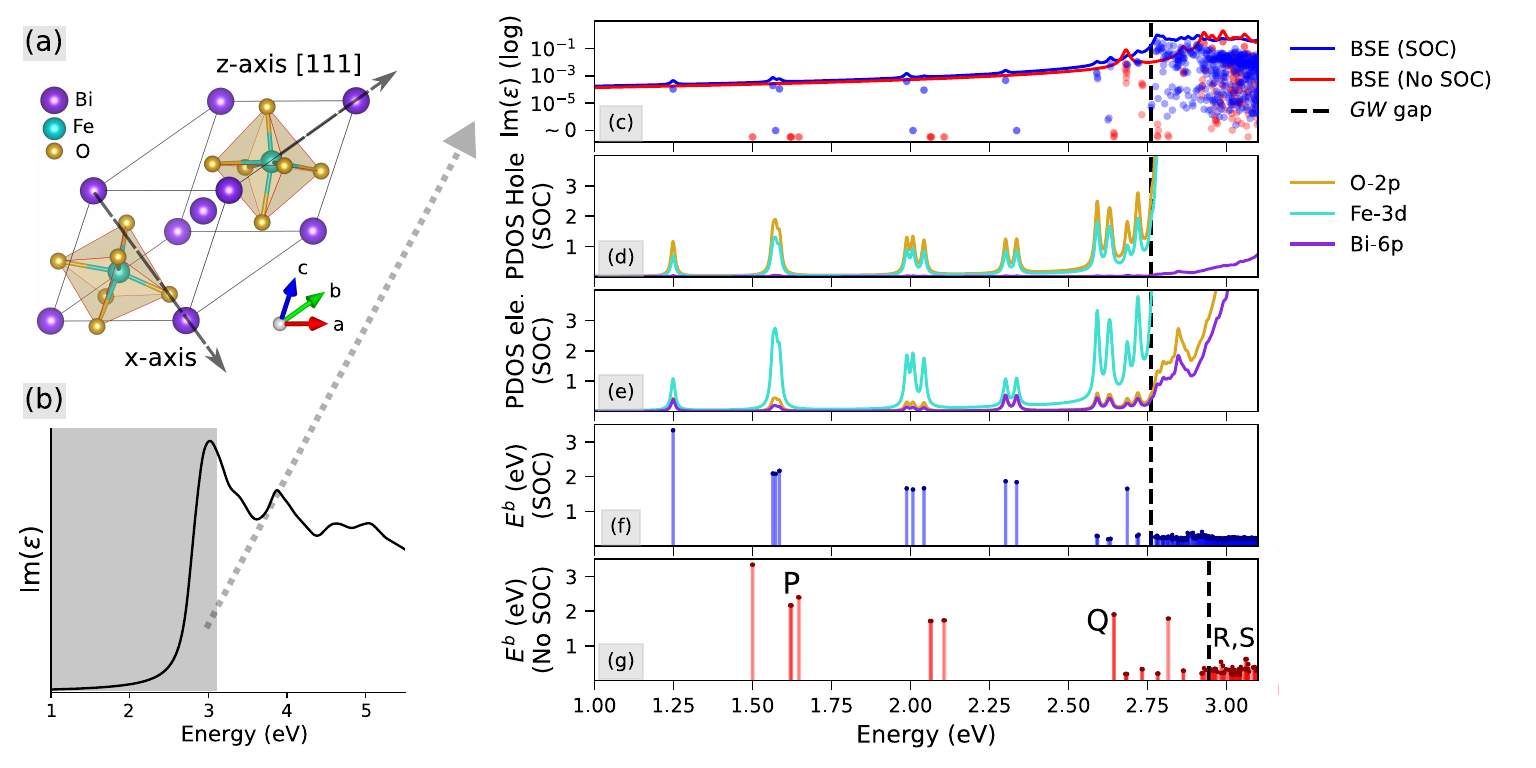}
    \caption{(a) Primitive unit cell of BiFeO$_3$.
    (b) Calculated Im($\varepsilon$) of BiFeO$_3$ using\textit{GW}-BSE. A uniform electronic broadening of 100~meV was used.
    (c) Im($\varepsilon$) of BiFeO$_3$, in the low energy range, corresponding to light polarized perpendicularly to the z-axis. An uniform electronic broadening of 7.5~meV is used to highlight the position of the excitons. The dots correspond to the exciton energies and oscillator strengths. Blue (red) lines and dots correspond to the case with SOC (without SOC).
    (d,e) Exciton PDOS weighted with overlap of electron and hole states with atomic orbitals.
    (f,g) Exciton binding energies with and without consideration of SOC, respectively.}
    \label{fig:bse_onset}
\end{figure*}

Fig.~\ref{fig:bse_onset}a shows the primitive unit cell of BiFeO$_3$.
The z-axis is along the [111] crystal direction and is the principal axis of the underlying $C_{3\mathrm{v}}$ point group.
Fig.~\ref{fig:bse_onset}b depicts the calculated imaginary part of dielectric function Im($\varepsilon$) of BiFeO$_3$ for unpolarised light, averaged over all polarization vectors.
It reproduces the features and peak positions of the experimentally observed spectrum, as detailed in our companion work~\cite{aseem_sven_prb}.  In Fig.~\ref{fig:bse_onset}c, we focus on the low-energy part of the calculated Im($\varepsilon$) for light polarized in the x-y plane (lines) and the corresponding oscillator strengths (dots) of the lowest-energy excitons, both with (blue) and without (red) consideration of the SOC.
Among excitons which are well-separated from the absorption onset at 2.76 eV, the exciton spectrum features a pair of states at $\sim$1.25 eV, followed by sets of excitons at $\sim$1.56 eV, $\sim$2.0 eV, and $\sim$2.30 eV.
The existence of some of these in-gap excitations has been verified by experiment, where they appear as two distinct features with an intensity of three orders of magnitude smaller than that of the peaks above the band gap~\cite{xu_optical_2009,meggle_2019,ramirez_magnon_2009,ramachandran_charge_2010,gomez-salces_effect_2012}.  \textcolor{black}{The computed spectrum is compared with measured spectra from literature, within low energy range, in the supplemental material (SM)~\cite{si}, in the light of past works cited therein~\cite{gomez-salces_effect_2012,xu_optical_2009,meggle_temperature-dependent_2019,burns1993mineralogical,alejandro_magnons}.}

So far, these excitations have been described as excitations within the Fe $3d$-shell and labeled as $T_{2\mathrm{g}}$ transitions, in the language of crystal field theory for  octahedral transition metal coordination complexes ~\cite{ramachandran_charge_2010,xu_optical_2009,ramirez_magnon_2009,gomez-salces_effect_2012,meggle_2019,burns1993mineralogical}.
These transitions are strongly affected by SOC, which leads to both a sizable red-shifting of the excitons and to their finite oscillator strength, confirming their spin-flip nature. 
To further characterize these transitions, we analyze both the electron and hole in terms of their atomic orbital composition.
To this end, we define an orbital-projected density of states (PDOS) for the electron for each atomic orbital $|i\rangle$ as
\begin{equation}\label{eq:PDOES_bse}
  \text{PDOS}_{i}^{\text{elec}}(\omega) = \sum_{S}\sum_{\mathbf{k},c,v}  \left| A^S_{\mathbf{k},c,v} \right|^2
  \left| \langle i | \mathbf{k},c \rangle \right|^2 \delta(\hbar\omega - E_S),
\end{equation} 
and similarly for the hole.
The resulting PDOSs are shown in Fig.~\ref{fig:bse_onset}d and e.
They show that, for the in-gap excitons, the hole is composed of O-2$p$ and Fe-3$d$ hybridized orbitals, while the electron is composed predominantly of Fe-3$d$ orbitals, with sizable contributions from O-2$p$ and Bi-6$p$ orbitals in certain cases. 
This clarifies the mixed composition of these excitations and illustrates the importance of the spin-orbit coupling mediated by the Bi atoms. 

These excitons are tightly bound excitations, and we quantify this by calculating the negative of the expectation value of the electron-hole interaction in the excitonic state:
\begin{equation}\label{e_binding}
  E_{S}^{\mathrm{b}} \equiv - \langle S| \hat{K}_{\mathrm{el-h}} |S \rangle
  = \sum_{\mathbf{k},c,v} \left| A^S_{\mathbf{k},c,v} \right|^2
    \left( \varepsilon^{\mathrm{QP}}_{\mathbf{k},c} - \varepsilon^{\mathrm{QP}}_{\mathbf{k},v} \right) - E_S,
\end{equation}
where $\varepsilon^{\mathrm{QP}}_{\mathbf{k},c(v)}$ denotes the quasi-particle energies of the conduction (valence) band states.
The thus-defined binding energies $E_b$ are shown in Fig.~\ref{fig:bse_onset}f (with SOC) and g (without SOC).
While the electron-hole interaction kernel comprises of both the attractive screened Coulomb interaction and the repulsive exchange interaction, we note that the contribution of the latter is very small ($10^{-3}$-$10^{-4}$~eV).
The exceptionally high values of the binding energies of these excitons, which reach values of up to twice the total exciton energy, suggest a high degree of spatial localization. 
In total, there are 20 strongly bound excitons, which are discussed and explained in our companion paper~\cite{aseem_sven_prb} in terms of the crystal field splitting.
By contrast, for the excitons located in energy above the absorption edge, the binding energy is much lower ($\sim 0.15-0.3$~eV) and comparable to other bulk semiconductors with delocalized excitons.
The SOC further leads to a splitting of the exciton levels that ranges from $\sim8$~meV for the strongest bound exciton at 1.56~eV up to a splitting of 34 meV for the exciton at 2.30~eV.
This amount of splitting is consistent with the small amount of Bi-6$p$ orbital admixture in the electron part of the exciton.
Beyond this, the overall energy ordering and the general atomic orbital character of the strongly bound in-gap excitons are unaffected by the SOC.

\begin{figure*}[]
    \centering
    \includegraphics[width=0.85\textwidth]{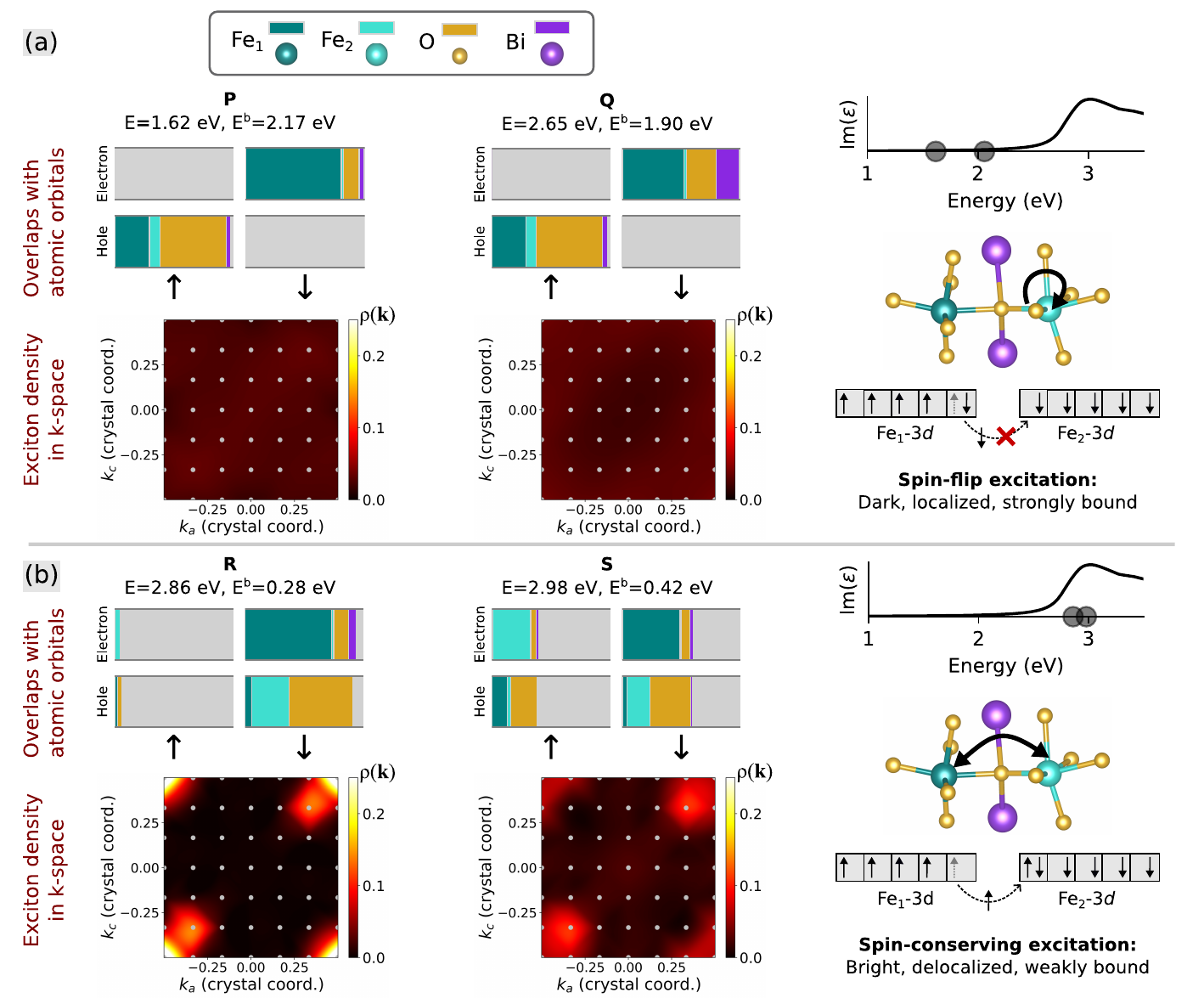}\caption{(a) Analysis of the spin-flip excitons P and Q in terms of their exciton density in the wave vector space and overlaps with atomic orbitals. Gray dots in the exciton density plots depict the discrete sampling of the wave vector space. 
    (b) Same as (a) for the spin-conserving excitons R and S. The location of the excitons on the energy axis are depicted in the adjacent plot of Im($\varepsilon$).}
    \label{fig:exciton-detail}
\end{figure*}

In Fig.~\ref{fig:exciton-detail}, we take a closer look at two examples each of the strongly bound in-gap (panel a) and the weakly bound above-gap excitons (panel b).
To arrive at a simpler approximate picture, we focus on the case without SOC which allows the resolution of the spins of the bound electron-hole pair.
For the strongly bound in-gap excitons in Fig.~\ref{fig:exciton-detail}a (labeled P and Q in Fig.~\ref{fig:bse_onset}g), the spin-resolved excitonic overlaps of the electron and hole with atomic orbitals show that they have opposite spins and hence constitute spin-flip excitations.
We can analyze them further by separating the contributions of the 3$d$ orbitals of each of the two Fe atoms in the unit cell (dark and light shades of teal).
We find that the Fe content of both the electron and hole states is mostly localized on the \emph{same} Fe atom.
A heat map of the exciton density in wave vector space,
\begin{equation}
    \rho^{S}(\mathbf{k}) = \sum_{c,v}  \left| A^S_{\mathbf{k},c,v} \right|^2,
\end{equation}
integrated along the [010] crystal direction, further confirms that these excitons are completely delocalized in reciprocal space, \textit{i.e.}, very localized in real space.

This unambiguously confirms the existing speculation that these in-gap excitations are to a large part on-site Fe transitions in which the electron spin is flipped.
Moreover, the local nature of the excitation explains the excessively high binding energy, stemming from $1/r$ behavior of the Coulomb interaction and a weak dielectric screening of this interaction at short length-scales by other on-site electrons.
Lastly, we note that the approximate anti-ferromagnetic spin structure necessitates a spin-flip for on-site intra-$d$-shell transitions owing to the $^{5}d$ half filling, which prevents the hopping of a flipped spin from one Fe atom to a neighboring one due to the Pauli exclusion principle (see sketch in Fig 2a \textcolor{black}{and Fig. S3 of the SM for real-space plots of excitonic electron and hole}). 

By contrast, the excitons above the gap show the opposite behavior, see Fig.~\ref{fig:exciton-detail} (marked as R and S in Fig.~\ref{fig:bse_onset}g).
The spin composition of the hole and the electron for these excitons is identical, \textit{i.e.}, they do not involve a flipping of the spin.
In addition, the Fe contributions of hole and electron stem from two \emph{different}, neighboring Fe atoms.
This delocalization of the exciton is further confirmed by its distribution in wave vector space, where it is localized in a specific corner of the first Brillouin zone, \textit{i.e.}, in real space it is delocalized.
This suggests that the above-gap excitations are of a very different type than the in-gap excitations and mostly involve the hopping of an electron from one Fe atom to one of its neighbors while keeping its spin orientation, in non-violation of the Pauli principle.
The delocalization of this type of exciton also explains their much lower binding energies compared to the strongly localized in-gap excitations.

\begin{figure*}[]
    \centering
    \includegraphics[width=0.85\textwidth]{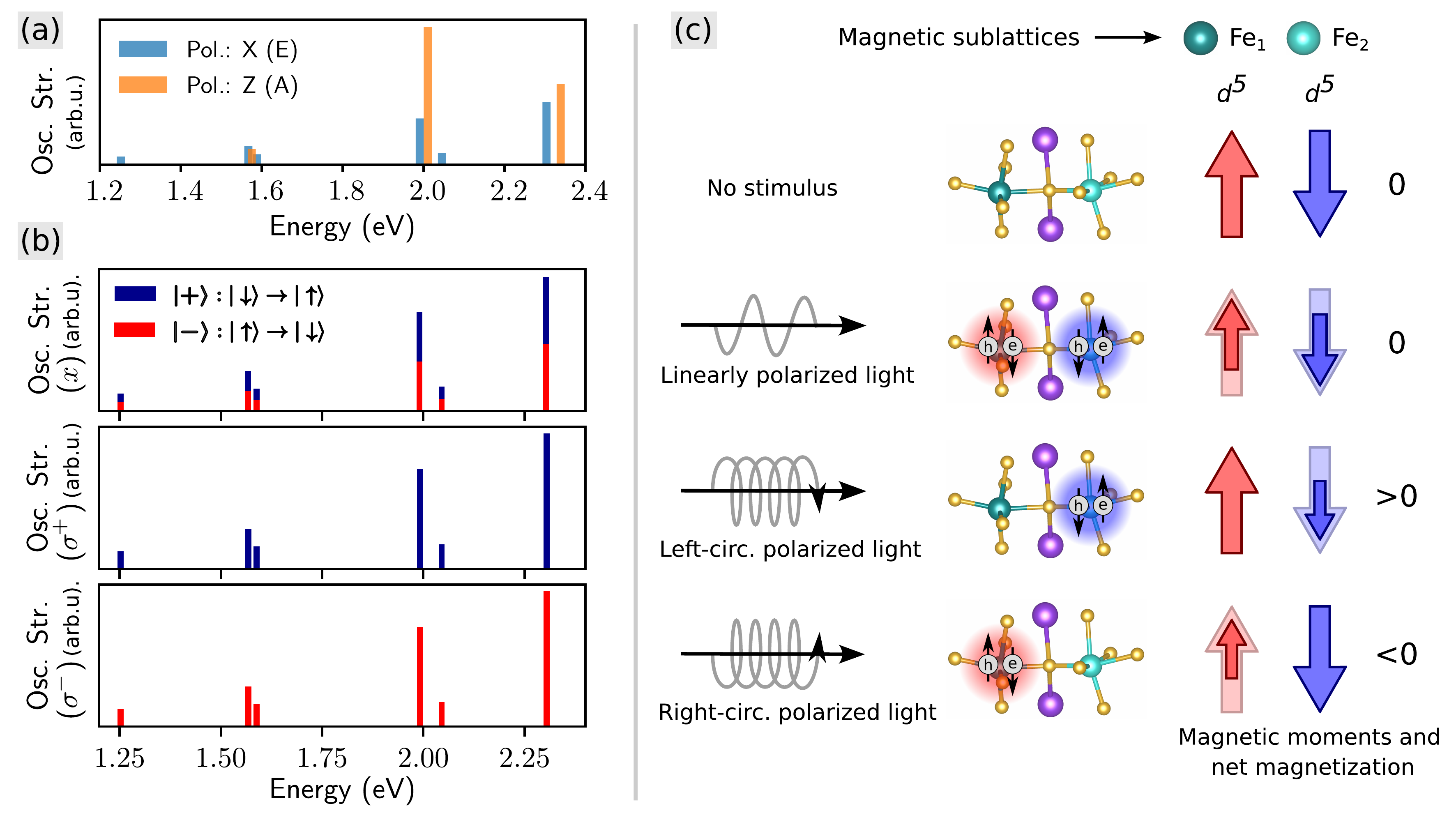}
    \caption{(a) Oscillator strengths of the lowest-energy bound excitons in BiFeO$_3$ for light polarized along the $x$- (blue bars) and $z$-direction (orange bars).
    (b) Oscillator strengths of doubly degenerate, spin-polarized ``E''-excitons for linearly (x-) and circularly ($\sigma^{\pm}$-)polarized light. The oscillator strengths are decomposed into contributions from excitons localized on Fe centers with opposite spin magnetization and, thereby, corresponding to $\lvert \uparrow \rangle \rightarrow \lvert \downarrow \rangle$ (red bars) and $\lvert \downarrow \rangle \rightarrow \lvert \uparrow \rangle$ (blue bars) spin flips, on the Fe$_1$ and Fe$_2$ sites, respectively.
    (c) Illustration of the light-induced spin-magnetization in BiFeO$_3$ upon excitation of an ``E'' exciton using linearly and circularly polarized light.}
    \label{fig:circ_pol}
\end{figure*}

We can exploit the quantum mechanical selection rules to specifically target the localized spin-flip excitations to create a localized spin excitation using light of a certain polarization.
For this, we observe that the space group of BiFeO$_3$, $R3c$, features three irreducible representations, $A_1$, $A_2$, and $E$, according to which all zero-momentum excitations can be classified.
The $E$~excitons correspond to a linear combination of states that carry an angular momentum of $\pm \hbar$ along the z-axis and hence can couple to light propagating along the $z$-axis, which is polarized in the $x$-$y$-plane.
In contrast, the $A_1$- and $A_2$-excitons possess zero angular momentum along the $z$-direction and only couple to light polarized along the $z$-axis.
Based on group theory alone, we would thus expect an excitation spectrum consisting of doubly degenerate ($E$) and singly degenerate ($A_1$, $A_2$) excitons, with all degenerate excitons coupling with light polarized within the $x$-$y$ plane and all non-degenerate excitons coupling with light polarized along the z-direction.
While the results of our calculation follow these selection rules, the lowest-lying, strongly bound $A_1$- and $A_2$-excitons are almost degenerate nonetheless, while possessing sizably different oscillator strengths.
This quasi-degeneracy can be attributed to the strong binding between the electron and hole, which localizes the exciton on one of the two Fe sites, where the two excitons acquire opposite spin-excitation character.
Due to the spin-conserving nature of the Coulomb interaction, these two localized excitations can only couple through the very weak exchange interaction and hence remain quasi-degenerate.

In Fig.~\ref{fig:circ_pol}a, we show the decomposition of the exciton spectrum into $A$- and $E$-excitons, which we distinguish by their optical activity in or perpendicular to the $x$-$y$-plane.
The two members of each $E$-exciton doublet transform in opposite ways under 120$^\circ$ rotations, picking up a phase factor of $\exp(\pm i 2\pi/3)$, and are mixed by the rototranslation elements of the $R3c$ group along the [111] skew axis, which effectively exchange the two Fe sites.
This implies that one can choose a basis in which the two members of an $E$ doublet carry a definite angular momentum and that the admixtures of orbitals from the two distinct Fe atoms contributing to each member are exactly opposite.
It further implies that, if an excitation is localized on only one of the Fe sites, it necessarily comes with a partner excitation localized on the other Fe atom of the same energy and that both of these excitations carry opposite angular momentum.
This is precisely the case for the lowest lying $E$ excitons, which are strongly bound by the electron-hole Coulomb attraction and are consequently localized on one Fe site only, with the coupling between excitons on different Fe sites being negligible (see Fig. S3).
From the above, it then follows that group theory predicts these particular $E$ exciton doublets to carry a definite angular momentum, i.e., the $E$ doublet can be chosen to be both site- and angular momentum-diagonal. 
We can understand this further by noting that in the absence of a strong coupling between excitons localized on different Fe sites, the space group of the crystal can effectively be replaced by two individual $C_3$ point groups for each Fe site separately.
As $C_3$ only allows one-dimensional representations, localized excitations are thus guaranteed to have definite angular momentum, while the actual exact symmetry of the crystal ensures that the two sets of excitations on the different Fe atoms have opposite angular momentum. 
These pairs of localized excitons with definite angular momentum then couple selectively to either left- or right-circularly polarized light and can thus be considered \emph{chiral} excitons.
This selective coupling can be exploited to determine the angular momentum content of the exciton states and to obtain angular momentum \textit{pure} states. The exciton doublet calculated using BSE ($\lvert \alpha \rangle, \lvert \beta \rangle$) may not necessarily correspond to \textit{pure} states.  However, rotating these states within the degenerate subspace to $\lvert + \rangle, \lvert - \rangle$ states such that they couple exclusively with either left- or right-circularly polarized light ensures that $\lvert + \rangle, \lvert - \rangle$ are \textit{pure} states with angular momentum $+ \hbar$ and $- \hbar$, respectively. 

Despite the strong spin-orbit coupling present in BiFeO$_3$, these strongly bound $E$ exciton pairs are also very pure in terms of the spin angular momentum they carry.
To demonstrate this, we calculate the expectation value of the $z$-component of the total spin operator in the exciton state $|S\rangle$ from the expression
\begin{equation}
\begin{split}
    \langle S| \hat{S}_{z} |S \rangle = \sum_{\mathbf{k},c,v}
    \Big[ &
    \sum_{c'} \left(A^S_{\mathbf{k},c',v}\right)^*
    \langle \mathbf{k},c'| \hat{S}_z |\mathbf{k},c\rangle
    A^S_{\mathbf{k},c,v} \\
    & - \sum_{v'} \left(A^S_{\mathbf{k},c,v'}\right)^*
    \langle \mathbf{k},v| \hat{S}_z |\mathbf{k},v'\rangle
    A^S_{\mathbf{k},c,v}
    \Big],
\end{split}
\end{equation}
where the sums over $c,c'$ ($v,v'$) run over conduction (valence) band states, the matrix elements involving the Kohn-Sham states are calculated in a plane-wave-spinor basis.
We find that despite the strong spin-orbit interaction in BiFeO$_3$, the lowest bound $E$ excitons ($\lvert + \rangle, \lvert - \rangle$) possess a spin polarization of up to 97\%, i.e., their spin content is approximately all of the total angular momentum of $\pm \hbar$. \par 

As shown in Fig.~\ref{fig:circ_pol}b, linearly polarized light along the x-axis excites both $\lvert + \rangle$ and $\lvert - \rangle$ exciton states localized on Fe$_1$ and Fe$_2$ respectively. However, using right- ($\sigma^+$) or left- ($\sigma^-$) CPL creates an excited state with definite angular momentum and that corresponds to a specific spin-flip, localized exclusively on Fe$_1$ or Fe$_2$. 
Consequently, these excitations provide an excellent way to create a highly localized, highly spin-polarized excitation with a favorable radiative lifetime due to its low oscillator strength.
As they can be selectively stimulated by light of a fixed polarization, these chiral excitons are ideal candidates for quantum information applications. \par 

By targeting a specific spin-flip excitation using CPL, it is possible to induce a net spin magnetization in the normally antiferromagnetic BiFeO$_3$, transiently altering its local magnetic texture.
In Fig.~\ref{fig:circ_pol}c, we illustrate that the angular momentum selectivity of CPL allows alteration of the local magnetic moment associated with either of the two magnetic sublattices Fe$_1$ or Fe$_2$ by targeting the $\lvert - \rangle$ and $\lvert + \rangle$ $E$ excitons localized on these sublattices. \textcolor{black}{In the SM, we provide a back-of-the-envelope estimation of the rate of creation of CPL-induced magnetization, indicating that such magnetization would be in an observable realm. } \\
\textcolor{black}{The theoretical investigation presented here is also useful for interpreting observations regarding helicity-dependent optical response of $d-d$ transitions in other transition metal compounds. 
Recently, in case of antiferromagnetic Fe$_2$Mo$_3$O$_8$, time-resolved measurements of magneto-optical Kerr effect by Sheu et al. confirmed transient CPL-induced magnetization due to $d-d$ transitions\cite{Sheu2019-kk,Zhuang2023-hm}. Interestingly, the observations of CPL-induced magnetization are consistent with the symmetry rules predicted here relating direction of propagation of light with change in magnetic moment in the context of 120$^\circ$ rotational symmetry, featured also by Fe$_2$Mo$_3$O$_8$. The presented theoretical frame can also be used to understand the quantum mechanical origin of helical photoluminescence observed in van der Waals antiferromagnet CrI$_3$~\cite{Seyler2017,Grzeszczyk2023-ze}. } 

In conclusion, we have investigated the low-energy part of the electronic excitation spectrum of BiFeO$_3$ using state-of-the-art \textit{ab initio} and many-body perturbation theory methods.
We find that the lowest energy electronic excitations in BiFeO$_3$ correspond to strongly bound, spin-flip excitations that are localized on one particular Fe site.
These excitations are chiral and possess an almost pure spin content that can selectively be excited with left- and right- circularly polarized light.
This makes these excitations\textcolor{black}{, present in a wide range of transition metal compounds,} very interesting for spintronic and magneto-optical applications, as they can serve to allow a manipulation of the spin degree of freedom and of the magnetization with light. \par 

\begin{acknowledgments}

We acknowledge funding from the National Research Fund (FNR) Luxembourg, project “RESRAMAN” (grant no. C20/MS/14802965).
The \textit{ab initio} calculations are performed on the University of Luxembourg's high-performance computing (ULHPC) clusters~\cite{Varrette2022-td}.
We thank colleagues of Theoretical Solid State Physics group at University of Luxembourg, Muralidhar Nalabothula and Ludger Wirtz, for valuable input and fruitful discussions. We also acknowledge SWIPE project partners, in particular, Georgy Gordeev, for scientific exchanges. 

\end{acknowledgments}

\bibliography{apssamp}
\bibliographystyle{apsrev4-2}

\end{document}


\maketitle

\noindent\rule{\textwidth}{0.5pt}

\tableofcontents

\noindent\rule{\textwidth}{0.5pt}
\vspace{0.25cm}



\section{Optical response of $d-d$ excitations }

\begin{figure}[h!]
    \centering
    \includegraphics[width=0.75\textwidth]{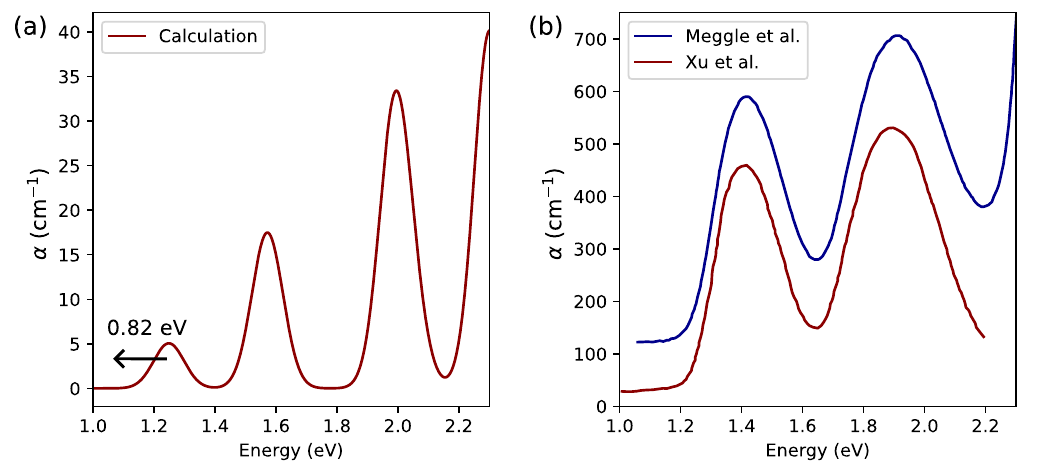}
    \caption{Low energy absorption spectra of BiFeO$_3$:  (a) calculated using BSE/$evGW$ approach and (b) experimentally measured, taken from ~\cite{meggle_temperature_2019} and ~\cite{xu_optical_2009}. As discussed in the manuscript, the peak at 1.25 eV in the calculated spectra (a) corresponds to unconverged magnon states. Gaussian broadening of 50 meV is used for the calculated optical spectra.  }
    \label{fig:alpha}
\end{figure}

Fig.~\ref{fig:alpha} depicts the low-energy optical spectra of BiFeO$_3$ calculated by solving the Bethe-Salpeter equation (BSE) as well as experimentally measured optical spectra from literature. We note that our calculations underestimate the optical absorption due to $d-d$ transitions by an order of magnitude. This can be attributed to several factors, including enhancement of the absorption features due to coupling of exciton states with lattice vibrations (evident from significant broadening of the peaks)~\cite{gomez-salces_effect_2012,burns1993mineralogical}, an overlap of the tail of bright excitations at higher energy, as well as inadequacies of the semilocal DFT functional employed for calculation of the single-electron wavefunctions, which trickle down in dipole matrix elements.  In the calculation of the optical spectra in Fig. ~\ref{fig:alpha} (a), we consider that light polarization is perpendicular to the ferroelectric axis.  However, the direction of polarization of light is unknown in the case of the measured optical spectra reproduced in Fig.~\ref{fig:alpha} (b) from ref.~\cite{xu_optical_2009,meggle_temperature_2019}. We further note that the lowest energy peak corresponds to unconverged states (extrapolated energy: 0.82 eV). These unconverged states are probably magnonic excitations which are captured by BSE~\cite{alejandro_magnons,Olsen2021-fi}. 

\section{Estimation of rate of magnetization}

Let us assume that crystalline BiFeO$_3$ is irradiated with monochromatic light of frequency $\omega$. The rate of absorption of photons per unit time is given by the Fermi's golden rule: 

\begin{equation}
    \Gamma (\omega) = \dfrac{2\pi}{\hbar}  \, \sum_S \, \lvert    \langle S | H^{\prime} | 0 \rangle \rvert^2 \, \delta(\hbar\omega-\Omega_S)
\end{equation}

where $H^{\prime} $ is the interaction of exciton ($\lvert S \rangle$) with the radiation electric field, and $\Omega_S$ is energy of exciton. Within the dipole approximation, it is given as 
\begin{equation}
 H^{\prime} =  -e\, \dfrac{E_0 (\omega) }{2} \, \mathbf{\hat{r}.\hat{e}}
\end{equation}

where $-e$ is the charge of an electron, $\mathbf{\hat{r}}$ is the position operator, $E_0 (\omega)$ is the amplitude of the electric field, and $\mathbf{\hat{e}}$ is the unit vector along the polarization of the electric field. The factor of 2 accounts for consideration of only the absorption process (excluding stimulated emission). \par 
The transition probability depends on the dipole matrix elements ($D^S_{\mathbf{e}}$) corresponding to coupling between the  exciton state $\lvert S \rangle$ and electric field polarization along $\mathbf{\hat{e}}$, written as
\begin{equation}
D^S_{\mathbf{e}} \, = \, \langle S|\mathbf{\hat{r}.\hat{e}}|0\rangle = \left( \,\sum_{\mathbf{k},c,v} \left(A^S_{\mathbf{k},c,v}\right)^* \langle\mathbf{k},c|\mathbf{\hat{r}}|\mathbf{k},v\rangle \, \right) . \mathbf{\hat{e}}
\end{equation}

where  $A^S_{\mathbf{k},c,v}$ is the exciton wavefunction in the basis of transitions from valence ($\mathbf{k},v$) to conduction bands ($\mathbf{k},c$), obtained by solving Bethe-Salpeter equation (BSE).  Then the absorption rate can be written as 
\begin{equation}
    \Gamma (\omega) = \dfrac{\pi e^2 E_0^2}{2\hbar N_{\mathbf{k}} V_{uc}}  \, \sum_S \, \lvert     D^S_{\mathbf{e}} \rvert^2 \, \delta(\hbar\omega-\Omega_S)
\end{equation}

where $V_{uc}$ is the volume of unit cell and $N_k$ is the number of k-points sampled.

Using the relation between $E_0$ and the intensity of light ($I$)  \textit{viz}. $E_0^2 = (2/ n_r c \epsilon_0) I$ ($n_r$ : refractive index), the rate of absorption can be reformulated. 

\begin{equation}
    \Gamma (\omega) = \dfrac{\pi e^2 I}{\hbar n_r c \epsilon_0 N_{\mathbf{k}} V_{uc}}  \, \sum_S \, \lvert     D^S_{\mathbf{e}} \rvert^2 \, \delta(\hbar\omega-\Omega_S)
\end{equation}

\begin{equation}
    \Gamma (\omega) = \dfrac{\epsilon_i(\omega) I}{\hbar n_r c}  = \dfrac{I \alpha(\omega)}{\hbar \omega}
\end{equation}
Here, $\epsilon_i(\omega)$ is the imaginary part of the dielectric function and $\alpha(\omega)$ is the absorption coefficient.

Let us now concern ourselves with the excitation of the $E$ exciton doublets that lie deep within the band gap and form chiral pairs of quantum states coupling exclusively with left and right circularly polarized light. Henceforth, we only consider circularly polarized light with polarization $\mathbf{\hat{e}}$ perpendicular to the ferroelectric axis of the crystal ($z$). As discussed in the manuscript, these chiral excitations bring about flipping of electron spin, thereby amounting to a net change in the spin magnetic moment $\approx \pm 2 \mu_{_B} $  corresponding to the circular polarization of light ($+$ : left and $-$: right). With this, the rate of magnetization can be written as

\begin{equation}
\Gamma_{\text{mag}} (\omega) = \pm \dfrac{2\pi e^2 \mu_{_B} I}{\hbar c \epsilon_0 N_{\mathbf{k}} V_{uc}}  \, \sum_S \, \lvert     D^S_{\pm} \rvert^2 \, \delta(\hbar\omega-\Omega_S) = \pm  \dfrac{2 \mu_{_B} I \alpha(\omega)}{\hbar \omega} 
\end{equation}

 Let us define rate of creation of magnetization per unit time and unit intensity of light as

\begin{equation}
R_{\text{mag}} (\omega) = \pm \dfrac{2\pi e^2 \mu_{_B}}{\hbar c \epsilon_0 N_{\mathbf{k}} V_{uc}}  \, \sum_S \, \lvert     D^S_{\pm} \rvert^2 \, \delta(\hbar\omega-\Omega_S) = \pm \dfrac{2 \mu_{_B} \alpha(\omega)}{\hbar  \omega} = \pm \dfrac{2 \mu_{_B} \epsilon_i(\omega)}{\hbar n_r c}
\end{equation} 

The prefactor can be evaluated in SI or other intuitive units. 

\begin{equation}
R_{\text{mag}} (\omega) = \pm \,586.6792055 \times \dfrac{\epsilon_i(\omega)}{n_r} \quad \text{A m$^{-1}$ . s$^{-1}$ .  W$^{-1}$ m$^{2}$ \, \text{(in SI units)}}
\end{equation} 

\begin{equation}
R_{\text{mag}} (\omega) = \pm \,6.32605745 \times 10^7 \times \dfrac{\epsilon_i(\omega)}{n_r} \quad \text{$\mu_{_B}$ cm$^{-3}$ . ps$^{-1}$ .  W$^{-1}$ m$^{2}$ \, }
\end{equation} 

\begin{figure}[h!]
    \centering
    \includegraphics[width=0.75\textwidth]{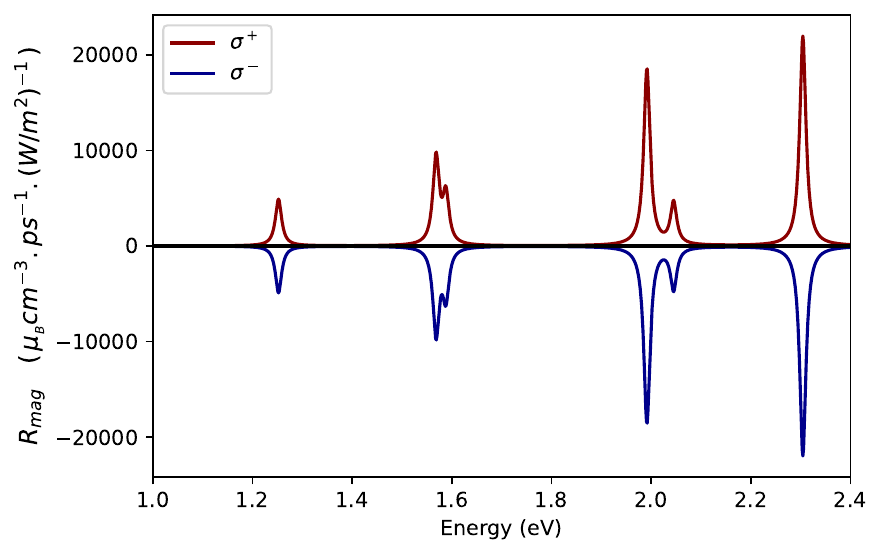}
    \caption{Rate of creation of light induced magnetization of BiFeO$_3$ using right ($\sigma^+$) or left ($\sigma^-$) circularly polarized light.}
    \label{fig:r_mag}
\end{figure}

\section{Real-space conditional density of excitons}

\begin{figure}[h!]
    \centering
    \includegraphics[width=0.98\textwidth]{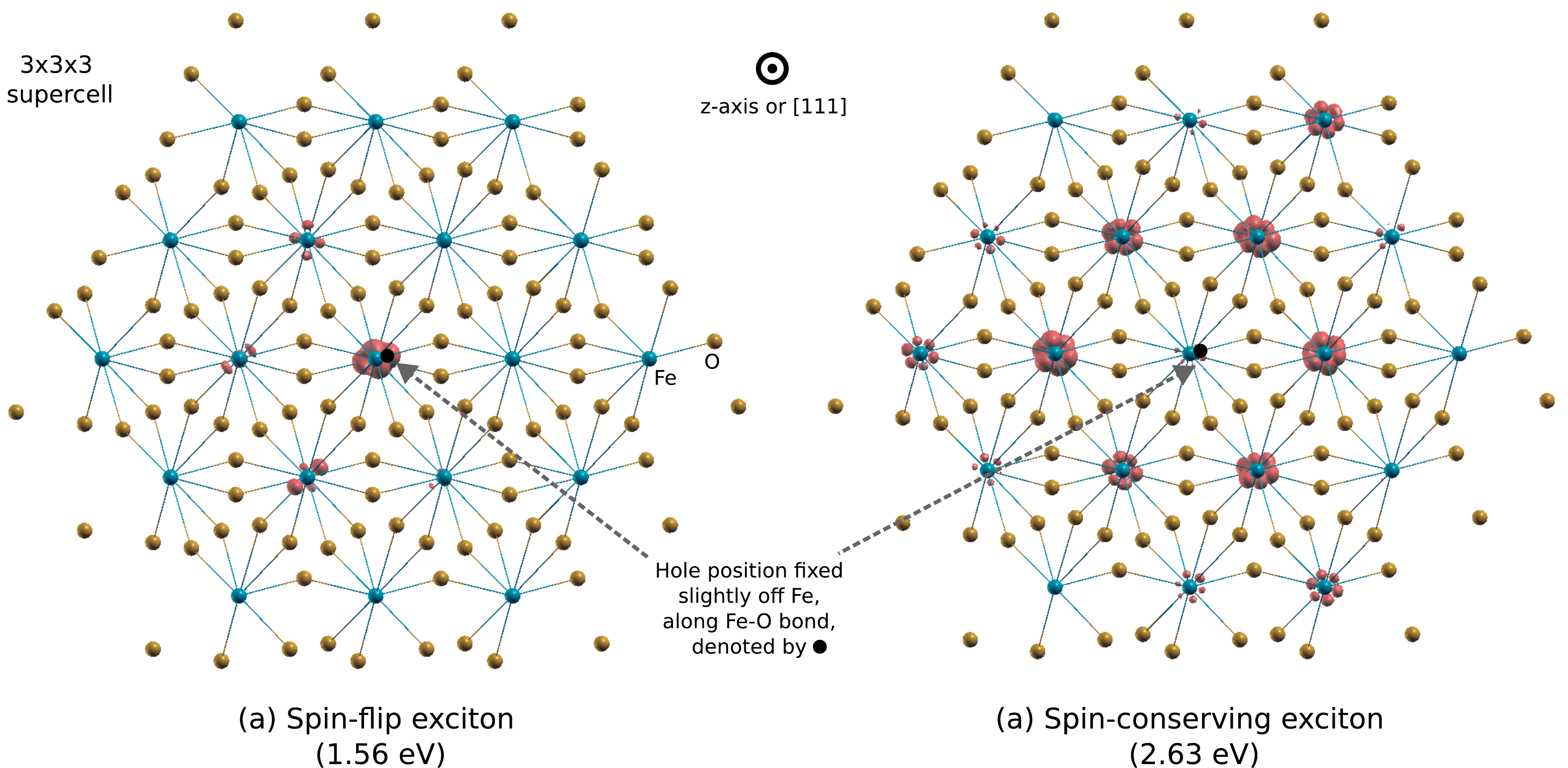}
    \caption{Real-space plots of conditional density of excitonic electron with fixed hole position.}
    \label{fig:exciton_wf}
\end{figure}


\FloatBarrier

\begingroup\footnotesize
\let\section\subsubsection
\makeatletter
\renewcommand\@openbib@code{\itemsep\z@}
\makeatother
\bibliography{refs_sm}
\bibliographystyle{nature}
\endgroup